\documentclass[twocolumn,preprintnumbers,superscriptaddress,nofootinbib,aps,prd,floatfix]{revtex4}

\usepackage{enumerate}
\usepackage{amsmath,amssymb}
\usepackage{graphicx}
\usepackage{bbm,slashed}
\usepackage{bbold}
\usepackage{xspace,slashed}
\usepackage{hyperref}
\hypersetup{colorlinks=true, citecolor=blue, urlcolor=blue, linkcolor=blue}
\usepackage[normalem]{ulem}
\usepackage{subfigure}

\usepackage{tabularx,booktabs} 
\newcolumntype{C}{>{$}c<{$}}
\AtBeginDocument{%
  \heavyrulewidth=.08em
  \lightrulewidth=.05em
  \cmidrulewidth=.03em
  \belowrulesep=.65ex
  \belowbottomsep=0pt
  \aboverulesep=.4ex
  \abovetopsep=0pt
  \cmidrulesep=\doublerulesep
  \cmidrulekern=.5em
  \defaultaddspace=.5em
}

\renewcommand{\vec}[1]{{\bf{#1}}}

\newcommand{\tx}[1]{{#1}}

\newcommand{\msbar}{\overline{\text{MS}}}

\begin{document}

\providecommand{\abs}[1]{\lvert#1\rvert}

\newcommand{\Znunujets}{(Z\to{\nu\bar{\nu}})+\text{jets}}
\newcommand{\Welnujets}{(W\to{\ell\nu})+\text{jets}}
\newcommand{\Znunujet}{(Z\to{\nu\bar{\nu}})+\text{jet}}
\newcommand{\Welnujet}{(W\to{\ell\nu})+\text{jet}}

\title{Implications of the muon anomalous magnetic moment for the LHC and MUonE}
\begin{abstract}
We consider the anomalous magnetic moment of the muon $a_\mu$, which shows a significant deviation from the Standard Model expectation given the recent 
measurements at Fermilab and BNL. We focus on Standard Model Effective Field Theory (SMEFT) with the aim to identify avenues for the upcoming LHC runs and future experiments such as MUonE. To this end, we include radiative effects to $a_\mu$ in SMEFT to connect the muon anomaly to potentially interesting searches at the LHC, specifically Higgs decays into muon pairs and such decays with resolved photons. Our investigation shows that similar to results for concrete UV extensions of the Standard Model, the Fermilab/BNL result can indicate strong coupling within the EFT framework and $a_\mu$ is increasingly sensitive to a single operator direction for high scale UV completions. In such cases, there is some complementarity between expected future experimental improvements, yet with considerable statistical challenges to match the precision provided by the recent $a_\mu$ measurement.
\end{abstract}

\author{Akanksha Bhardwaj}\email{akanksha.bhardwaj@glasgow.ac.uk} 
\affiliation{School of Physics \& Astronomy, University of Glasgow, Glasgow G12 8QQ, UK\\[0.1cm]}
\author{Christoph Englert} \email{christoph.englert@glasgow.ac.uk}
\affiliation{School of Physics \& Astronomy, University of Glasgow, Glasgow G12 8QQ, UK\\[0.1cm]}
\author{Panagiotis Stylianou}\email{p.stylianou.1@research.gla.ac.uk} 
\affiliation{School of Physics \& Astronomy, University of Glasgow, Glasgow G12 8QQ, UK\\[0.1cm]}

\maketitle

\section{Introduction}
\label{sec:intro}
The search for new physics beyond the Standard Model (SM), which so far has been unsuccessful, is one of the highest priorities of the current particle physics programme. A relevant anomaly in this context might be the recent measurement of the anomalous muon magnetic moment $a_\mu=(g-2)_\mu/2$ at Fermilab \cite{Muong-2:2021ojo}, which confirmed the earlier BNL E821 results~\cite{Muong-2:2004fok}, leading to a  $\sim 4\sigma$ tension~\cite{Aoyama:2012wk,Aoyama:2019ryr,Czarnecki:2002nt,Gnendiger:2013pva,Davier:2017zfy,Keshavarzi:2018mgv,Colangelo:2018mtw,Hoferichter:2019mqg,Davier:2019can,Keshavarzi:2019abf,Kurz:2014wya,Melnikov:2003xd,Masjuan:2017tvw,Colangelo:2017fiz,Hoferichter:2018kwz,Gerardin:2019vio,Bijnens:2019ghy,Colangelo:2019uex,Blum:2019ugy,Colangelo:2014qya} (see also~\cite{Aoyama:2020ynm} for a summary) %
\begin{equation}\label{eq:1}
\Delta a_\mu = a_\mu(\text{exp}) - a_\mu(\text{\text{SM}}) = (25.1 \pm 5.9) \times 10^{-10}\,.
\end{equation}
Anomalous magnetic moments are characterised by dimension-six operators $\sim \bar \psi_L \sigma^{\mu\nu} \psi_R F_{\mu\nu} + \text{h.c.}$ where $F$ denotes the QED field strength tensor.
Therefore, $a_\mu$ is directly sensitive to new interactions in renormalisable field theories and does not probe any of these theories' defining parameters. This is part of the reason why $a_\mu$ has received enormous attention from the BSM community, as it provides a formidably precise tool to constrain the structure of concrete UV extensions of the SM (see, e.g.~\cite{Athron:2021iuf} for a recent overview).

In contrast to these model-specific analyses, the null results of a plethora of new physics analyses at, e.g., the Large Hadron Collider (LHC) have highlighted model-independent methods based on effective field theory (EFT)~\cite{Weinberg:1978kz} techniques as alternative approaches to search for new physics effects. EFT is a powerful tool when extra degrees of freedom can be consistently integrated out~\cite{Buchalla:2013rka,Buchalla:2016bse,Cata:2015lta,Carmona:2021xtq,Dittmaier:2021fls} and when measurements are performed at energy scales that do not violate the scale hierarchies that are implicitly assumed by the EFT approach~\cite{Contino:2016jqw}. The latter can be challenging at hadron colliders with their large energy coverage and significant uncertainties~\cite{Englert:2014cva,Englert:2019rga}. The extraction of $a_\mu$ from data is largely free of such shortfalls and there has been a range of EFT-based investigations into the $a_\mu$ anomaly~\cite{Crivellin:2013hpa,Allwicher:2021jkr,Aebischer:2021uvt,Cirigliano:2021peb}.

Historically, EFT measurements have played a crucial role in shaping the understanding of physics of the weak scale. A famous example is the muon's lifetime providing a measurement of the Fermi constant $G_F$, the cutoff of the low-energy effective theory of the weak interactions. A constraint on the cutoff gives rise to an upper limit on a more fundamental mass scale. The latter is an important pointer towards experimental signatures (based on the assumption of a well-behaved perturbative expansion). In the SM 
\begin{equation}
{m \over \Lambda} = \sqrt{G_F m_W^2}\simeq 0.27\ll 1
\end{equation}
measures the UV completing degrees of freedom of Fermi's theory in units of its cutoff. 

The Fermi theory shows that new physics could appear at relatively low scales compared to the cutoff. This is strong motivation to consider $a_\mu$ in EFT to clarify implications for energy scales above the muon mass: In~\cite{Allwicher:2021jkr} it was shown that the combination of unitarity constraints and scale evolution could push the new physics scale to very large values (an observation that is echoed in concrete UV extensions, see e.g.~\cite{Baer:2021aax,Athron:2021iuf,Frank:2021nkq,Ellis:2021zmg,Zhang:2021dgl,Jueid:2021avn,Altmannshofer:2021hfu,Chakraborti:2020vjp,Chakraborti:2021kkr,Chakraborti:2021dli,Chakraborti:2021squ}). This raises the question whether $a_\mu$ merely fixes one parameter of the EFT Lagrangian, perhaps with little phenomenological implications for UV physics. The comprehensive analysis of~\cite{Aebischer:2021uvt} has evaluated the anomalous magnetic moment in Standard Model Effective Field Theory (SMEFT) with higher-order effects by matching and evolving $g-2$ in Low Energy Effective Field Theory (LEFT), including flavour-violating contributions, obtaining results consistent with the previous study of Ref.~\cite{Crivellin:2013hpa} which included loop contributions from nondiagonal contributions of ${\cal{O}}_{l e}$ and the dimension-six dipole operator. 

In this work, we employ SMEFT~\cite{Grzadkowski:2010es} to revisit the EFT context of $a_\mu$ with the aim to correlate $a_\mu$ with measurements at the high-luminosity LHC and muon-specific future experiments such as MUonE~\cite{Abbiendi:2016xup} (see also Refs.~\cite{ColuccioLeskow:2016dox,Crivellin:2020tsz,Fajfer:2021cxa,Crivellin:2021rbq,Paradisi:2022vqp} for correlations of $ a_\mu$ with additional modes for new physics models). To this end, we investigate $ a_\mu$ at full one-loop order in SMEFT, building on the results of~\cite{Dedes:2017zog,Dedes:2019uzs}. Our findings show that as discussed by previous effective interpretations of the muon $g-2$~\cite{Aebischer:2021uvt,Crivellin:2013hpa}, the only operators that could provide explanations for the Fermilab measurement individually are the dipole operators and ${\cal{O}}_{l e}$. 
To scrutinise further these operators, we consider a priori sensitive processes, such as the $h \mu \mu$ signal strength and constraints that can be placed by the future MUonE experiment. However, none of these avenues can provide comparable restrictions on the span of the dipole Wilson Coefficients (WCs) when compared to the Fermilab measurement.

We organise this work as follows: Sec.~\ref{sec:gm2} provides a short overview of $ a_\mu$ in SMEFT to make this work self-contained. 
We explore the impact of the higher orders at different scales in Sec.~\ref{subsec:uvopes} and additionally discuss the possibility that $ a_\mu$ arises as a radiative correction in Sec.~\ref{subsec:finops}. Different avenues with the potential to tension the dipole operators are studied in Sec.~\ref{sec:tension}, focusing on the decay of the $Z$ boson, the $h \to \mu \mu \gamma$ channel at LHC and also the MUonE experiment. We conclude in Sec.~\ref{sec:conc}.

\section{$a_\mu$ in SMEFT}
\label{sec:gm2}
Neglecting contributions to the unphysical (longitudinal) anapole moment~\cite{Czyz:1987xx}, the vertex function for muon-vector boson interactions can be expanded as 
\begin{widetext}
	\vspace{-1.8cm}
\begin{equation}
\label{eq:vtxdecomp}
	  \hfill  \parbox[h][0.3\linewidth][c]{0.25\linewidth}{  \includegraphics[scale=0.8]{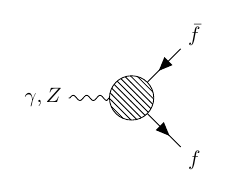} }  \hspace{-1.8cm}   \vspace{-1.8cm} 
	= - i e Q_f \bar{u}(p^\prime)  \bigg[ \gamma^\mu F_1(k^2) 
	+ i \frac{\sigma^{\mu\nu} k_\nu}{2 m_f} F_2(k^2) + \frac{\sigma^{\mu \nu} k_\nu}{2 m_f} \gamma_5 F_3(k^2) \bigg] u(p)\;,\hfill
\end{equation}
\end{widetext}
where the boson's momentum is given by $k = p^\prime - p$ and $\sigma_{\mu \nu} = i [ \gamma_\mu, \gamma_\nu] / 2$ are the usual Lorentz algebra generators. The fermion legs are on-shell and the electric charge $e$ is renormalised through the renormalisation condition $F_1(0) = 1$ at zero momentum transfer. The anomalous magnetic moment for the muon is then defined through the form factor as $a_\mu = F_2(0)$, while $F_3$ is related to the electric dipole moment.

The effective interactions of SMEFT in the Warsaw basis~\cite{Grzadkowski:2010es} that give rise to these form factors can be modelled with {\sc{SmeftFR}}~\cite{Dedes:2017zog,Dedes:2019uzs} which employs {\sc{FeynRules}}~\cite{Christensen:2008py,Alloul:2013bka} to obtain the relevant Feynman rules for the SMEFT Lagrangian truncated at operator-dimension six. Interfacing with {\sc{FeynArts}}~\cite{Hahn:2000kx} enables users to enumerate the relevant diagrams, and their respective amplitudes can be calculated with {\sc{FormCalc}}~\cite{Hahn:1998yk}. The tensor integrals that appear in these amplitudes are reduced to scalar Passarino-Veltman functions~\cite{Passarino:1978jh} (see also Ref.~\cite{Denner:2006fy}), which can then be evaluated analytically using {\sc{PackageX}}~\cite{Patel:2015tea}. The form factors can then be extracted from these expressions with the help of the Gordon identities which recast the result into the form of~Eq.~\eqref{eq:vtxdecomp}. For QED, this reduces to the well-known Schwinger result $a_\mu = \alpha/(2 \pi)$~\cite{Schwinger:1948iu}.

In contrast to the SM, tree level contributions to the anomalous magnetic moment are induced in SMEFT by the operators (see Refs.~\cite{Buchmuller:1985jz,Alonso:2013hga})
\begin{equation}
	\label{eq:amuops}
	\begin{split}
		{\cal{O}}_{e B}^{ f_1 f_2} =&\; (\bar{L}_{f_1} \sigma^{\mu \nu} e_{f_2}) \Phi\, B_{\mu \nu} \;,\\
		{\cal{O}}_{e W}^{ f_1 f_2} =&\; (\bar{L}_{f_1} \sigma^{\mu \nu}  e_{f_2}) \tau^I \Phi\,W^I_{\mu \nu}\;,
	\end{split}
\end{equation}
where $L$ ($e$) denotes the left-handed (right-handed) lepton, $B_{\mu \nu}$ and $W^I_{\mu \nu}$ are the field strength tensors for the $U(1)_Y$ and $SU(2)_L$ gauge groups, respectively, and $\Phi$ is the Higgs doublet. The Pauli matrices are denoted by $\tau^I$, $I=1,2,3$. The dimension-six operators induce an anomalous magnetic moment of\footnote{We consider only the diagonal entries of the operators in~Eq.~\eqref{eq:amuops} and suppress the flavour indices.}
\begin{equation}
	\label{eq:amutree}
	\Delta a_\mu^\tx{\text{tree}} = \frac{\sqrt{2} v m_\mu}{e \Lambda^2} \Big[ c_W \left(C_{e B} + C_{e B}^*\right) - s_W \left( C_{e W} + C_{e W}^* \right) \Big]\,,
\end{equation}
where $s_W$ ($c_W$) is the sine (cosine) of the Weinberg angle. Under the assumption that $C_{e W}$ and $C_{e B}$ are real, any contribution to the electric dipole moment is removed and the two operators only generate $\Delta a_\mu$ (modulo small SM electroweak radiative effects).

Extending the calculation to one-loop level requires additional renormalisation constants not present in the SM~\cite{Alonso:2013hga,Jenkins:2013wua,Jenkins:2013zja}. While no UV divergence is induced in the $F_2$ and $F_3$ form factors when only SM interactions are present, the additional SMEFT operators generate divergences that cannot be removed with the SM counterterms, but induce counterterms $\delta C_{e B}$ and $\delta C_{e W}$. By considering both the $\gamma \mu \mu$ and $Z \mu \mu$ vertices (alongside the renormalisation of $Z-\gamma$ mixing) we calculate the UV-divergent parts after dimensional regularisation (in dimensions $d = 4 - 2 \epsilon$) and subtract them in the $\msbar$ scheme such that both form factors are renormalised at one-loop order. 
To this end, we introduce the counterterm amplitude
\begin{widetext}
	\vspace{-1.8cm}
\begin{equation}
\label{eq:counter}
	\hfill\parbox[h][0.3\linewidth][c]{0.25\linewidth}{\includegraphics[scale=0.8]{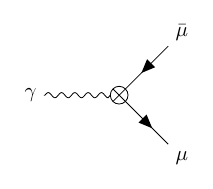}} \hspace{-1.8cm}\vspace{-1.8cm}
	= - i e Q_{\mu} \bar{u}(p^\prime)  \bigg[  
	i \frac{\sigma^{\mu\nu} k_\nu}{2 m_f} \delta F_2 + \frac{\sigma^{\mu \nu} k_\nu}{2 m_f} \gamma_5 ( i \delta F_3) \bigg] u(p)\;,\hfill
\end{equation}
\end{widetext}
where $\delta F_2$ and $\delta F_3$ renormalise the relevant physical Lorentz structures corresponding to $F_{2}(k^2)$ and $F_{3}(k^2)$, respectively. The factor of $i$ multiplying $\delta F_3$ is introduced just for convenience in expressing the equations later on. We do not consider divergences relevant to the $F_1$ form factor as this is renormalised in the Thomson limit to correspond to the correct electric charge~\cite{Denner:1991kt}. 
We express $\delta F_2$ and $\delta F_3$ in terms of the usual SM coupling and wavefunction renormalisation constants (see Ref.~\cite{Denner:1991kt}) and also introduce $\delta C_{e B}$ and $\delta C_{e W}$ as counterterms for the ${\cal{O}}_{e B}$ and ${\cal{O}}_{e W}$ interactions. This allows us to calculate 
\begin{widetext}
\begin{equation}
\begin{split}
	\delta F_{2,3} &= \frac{m_\mu}{\sqrt{2} e \Lambda^2} \bigg[ C_{e B} K_1 \pm C_{e B}^* K_1^\prime + C_{e W} K_2 \pm C_{e W}^* K_2^\prime + 2 c_W v ( \delta C_{e B} \pm \delta C_{e B}^* ) - 2 s_W v (\delta C_{e W} \pm \delta C_{e W}^*) \bigg] \;,\\	
	K_1 &= 2 c_W \delta v + c_W v ( \delta Z_{AA} + \delta Z_L^{\mu \mu,*} + \delta Z_R^{\mu \mu}) + v (2 \delta c_W - s_W \delta Z_{ZA})\;,\\
	K_2 &= - 2 s_W \delta v - s_W v ( \delta Z_{AA} + \delta Z_L^{\mu \mu,*} + \delta Z_R^{\mu \mu}) - v ( 2 \delta s_W  + c_W \delta Z_{ZA})\;,
\end{split}
	\label{eq:counterRCs}
\end{equation}
\end{widetext}
where the prime indicates $\delta Z_L^{\mu \mu, *} \to \delta Z_L^{\mu \mu}$ and $\delta Z_R^{\mu \mu} \to \delta Z_R^{\mu \mu,*}$. In the SM, the Higgs potential contains 
\begin{equation}
	V(\Phi) \supset v (\mu^2 + v^2 \lambda_H) h = t h\;,
\end{equation}
which is minimised at tree level via $t = 0$. Tadpole diagrams that capture the shift away from the classical Higgs field value due to higher orders are removed by introducing the counterterm $\delta t + t = 0$~(see Ref.~\cite{Denner:1991kt}). Expressing the vacuum expectation value as a function of the $W$ mass, Weinberg angle and electromagnetic coupling, we can formally identify
\begin{equation}
	\label{eq:renormvev}
	{\delta v\over v} =  \frac{\delta m_W^2}{2 m_W^2} + \frac{\delta s_W}{s_W} - \delta Z_e \;,
\end{equation}
which enters Eq.~\eqref{eq:counterRCs} ($\delta m_W^2$, $\delta s_W$ and $\delta Z_e$ are the counterterms of the W mass $m_W^2$, Weinberg sine angle $s_W$ and electric charge $e$ in the conventions of \cite{Denner:1991kt}). Alternatively, one can employ the so-called Fleischer-Jegerlehner scheme~\cite{Fleischer:1980ub} that shifts the bare vacuum expectation value by $\Delta v = - \delta t / m_H^2$, where $m_H$ is the Higgs mass, at the cost of inducing large corrections to renormalised quantities~\cite{Dekens:2019ept,Hartmann:2015oia,Cullen:2019nnr} (for discussions on this scheme see for example Refs.~\cite{Denner:2019vbn,Dudenas:2020ggt}).

As there are two independent operators contributing to $\Delta a_\mu$, we repeat the calculation for the $Z\mu\mu$ vertex which leads to similar counterterm expressions. Subsequently, we determine the values of $\delta C_{e B}$ and $\delta C_{e W}$ in the $\overline{\text{MS}}$ scheme by simultaneously requiring that the UV divergences in the $F_2$ and $F_3$ form factors for both the $\gamma \mu \mu$ and $Z \mu \mu$ vertices identically cancel and only finite terms for $d\to 4$ remain. 

The one-loop corrections to the magnetic moment in SMEFT also give rise to soft singularities at finite photon virtuality. These are soft (and universal) QED corrections to the dimension-six interactions and they cancel against soft photon emission off the dimension-six vertex of Eq.~\eqref{eq:vtxdecomp}. These soft singularities vanish in the limit of zero virtuality\footnote{We have checked this explicitly.}, reflecting the fact that the higher-dimensional operators are a manifestation of scale-suppressed new physics separated from universal soft (and collinear) effects in QED (see e.g.~\cite{Englert:2018byk} for a general discussion in the context of QCD). We can therefore omit soft singularities throughout this calculation.

Our calculation of $\Delta a_\mu$ is performed using $m_W$, $m_Z$ and the fine structure constant $\alpha$ as inputs of the theory.

\begin{table*}[t!]
\begin{tabular}{c||c|c}
\toprule
	WC (/$\Lambda^2$) & Fermilab/BNL allowed $[1/{\text{TeV}^2}]$ & SM allowed $[1/{\text{TeV}^2}]$ \\ 
\midrule
	$C_{e B}$ & $\left[8.21\times 10^{-6}, 1.33\times 10^{-5}\right]$ & $\left[-1.84\times 10^{-6}, 1.84\times 10^{-6}\right]$\\
	$C_{e W}$ & $\left[-1.41\times 10^{-5}, -2.27\times 10^{-5}\right]$ & $\left[3.15\times 10^{-6}, -3.15\times 10^{-6}\right]$\\
	$C_{W}$ & $\left[14.00, 22.60\right]$ & $\left[-3.14, 3.14\right]$\\
	$C_{\phi B}$ & $\left[0.36, 0.58\right]$ & $\left[-0.08, 0.08\right]$\\
	$C_{\phi W}$ & $\left[1.06, 1.71\right]$ & $\left[-0.24, 0.24\right]$\\
	$C_{\phi e}$ & $\left[7.92, 12.80\right]$ & $\left[-1.77, 1.77\right]$\\
	$C_{\phi l}^{(1)}$ & $\left[-8.27, -13.40\right]$ & $\left[1.85, -1.85\right]$\\
	$C_{\phi l}^{(3)}$ & $\left[-8.27, -13.40\right]$ & $\left[1.85, -1.85\right]$\\
	$C_{l e}$ & $\left[-1.69, -2.72\right]$ & $\left[0.38, -0.38\right]$
\\ \bottomrule
\end{tabular}
	\caption{Bounds on WCs in units of $\text{TeV}^{-2}$ for consistency with the Fermilab measurement and the errors of the SM prediction with $\mu = m_{\mu}$. We limit ourselves to WC operator values $\lesssim 4\pi/\text{TeV}^2$.}
	\label{tab:wcbounds}
\end{table*}

\medskip
In principle, semileptonic four-fermion operators of the $(\bar{L} R)(\bar{R} L)$ and $(\bar{L} R) (\bar{L} R)$ classes in the Warsaw basis~\cite{Grzadkowski:2010es} also contribute to the anomalous magnetic moment (see Ref.~\cite{Aebischer:2021uvt} for the generic case). However, we do not include them as such operators are often neglected by enforcing flavour symmetries, similar to the assumptions of Refs.~\cite{Aguilar-Saavedra:2018ksv,Ellis:2020unq} in the top sector. The structure of these operators can be generated from leptoquarks~\cite{Gherardi:2020det}, which have been explored extensively as explanations of the anomalous magnetic moment~\cite{Freitas:2022gqs,Chen:2022hle,Crivellin:2020tsz}.

Our aim is to identify operators that give rise to significant corrections to the SM that push the anomalous magnetic moment to larger values closer to the Fermilab/BNL result. As such we limit ourselves to operator directions that are not related to oblique electroweak precision constraints, i.e. we will neglect the ${\cal{O}}_{\phi W B}$ and ${\cal{O}}_{\phi D}$ operators due to their relations to the $S$ and $T$ parameters~\cite{Alonso:2013hga,Berthier:2015gja,Dedes:2017zog} 
(we note that the correlations of $a_\mu$ and electroweak data have been studied elsewhere, e.g.~\cite{Kanemitsu:2012dc,Cho:2000sf}). We have also considered only CP-conserving operators.
The remaining WCs are tabled in Tab.~\ref{tab:wcbounds} for a renormalisation scale choice of $\mu=m_\mu$ (the scale relevant to the $a_\mu$ measurement, we will discuss the impact of different choices, e.g., $\mu=\Lambda$ further below).\footnote{We have checked our results against the Renormalisation Group Evolution analysis of Ref.~\cite{Aebischer:2021uvt} and find very good agreement.} As can be seen, only a subset of these operator constraints is perturbatively meaningful. This reflects the general observation of the $a_\mu$ anomaly in theories with extended particle spectra~\cite{Athron:2021iuf,Anisha:2021jlz}.

\begin{figure}[!b]
	\includegraphics[width=0.45\textwidth]{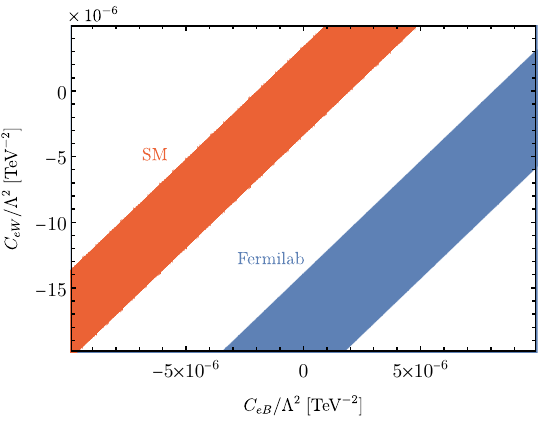}
	\caption{Bounds in the $C_{e W}$-$C_{e B}$ plane with the rest of the WCs fixed at zero from the anomalous magnetic moment with $\mu = m_\mu$.\label{fig:cewceb_amu}}
\end{figure}

\subsection{UV-divergent operators: $a_\mu$ as input parameter}
\label{subsec:uvopes}

\begin{figure}[!h]
	\includegraphics[width=8.7cm]{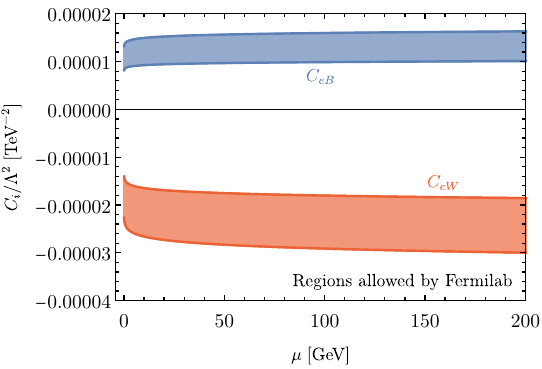}
	\caption{Individual intervals allowed by the Fermilab measurement for $C_{e B}$ and $C_{e W}$ (left) as a function of the dimensional regularisation scale $\mu$.
	\label{fig:mudependence}}
\end{figure}

\begin{figure}[!b]
	\includegraphics[width=0.45\textwidth]{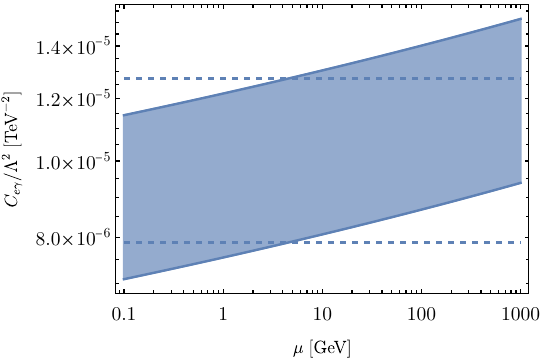}
	\caption{Individual interval allowed by the Fermilab measurement for $C_{e \gamma}$ including one-loop contributions is shown with solid lines and filled as a function of $\mu$. When only tree level contributions are included, the corresponding interval is shown with dashed lines.\label{fig:mudepcegamma}}
\end{figure}

From our included operators, the divergent parts of Eq.~\eqref{eq:vtxdecomp} depend only on $C_{e B}$, $C_{e W}$, $C_{\phi B}$ and $C_{\phi W}$. The latter two are constrained mainly from Simplified Template Cross Section (STXS) measurements and Higgs signal strength measurements at the LHC (see for example Refs.~\cite{Ellis:2020unq,Anisha:2021hgc}) without much potential to further clarify the muon anomalous magnetic moment measurement. The tree level $C_{e B}$ and $C_{e W}$ insertions on the other hand are hard to constrain due to chiral suppression. They are a subset of the few possible operators that can shift significantly the SM expectation when new physics appears at around the TeV scale. Contours including both operators are shown in Fig.~\ref{fig:cewceb_amu}. 

As any scheme- and scale-dependent parameter in QFT, the results understood in terms of Wilson coefficient constraints depend on the (unphysical) dimensional renormalisation scale $\mu$. From the point of matching calculations a choice of $\mu=\Lambda$ is intuitive, while scale-dependent logarithms are typically efficiently resummed by choosing an adapted renormalisation scale relative to the scale of measurement. This is familiar from many calculations in collider physics (e.g. scale choices of Higgs decays to bottom pairs~\cite{LHCHiggsCrossSectionWorkingGroup:2011wcg}). It is therefore worthwhile to investigate the $\mu$ scale choice as an indication of the reliability of our calculation with regards to neglected higher-order effects.
We show this in Fig.~\ref{fig:mudependence} for $\Lambda = 1$~TeV, highlighting that the effect on the WCs of dipole operators is essentially a shift of the allowed interval due to the presence of the tree level contribution. Considering the Fermilab measurement as input, the values of the WCs would need to increase as $\mu \to \Lambda$. In contrast, the dependence on $C_{\phi B}$ and $C_{\phi W}$ through logarithms $\log(m^2/\mu^2)$ (with $m$ representing a SM mass scale) leads to a different behaviour as a function of $\mu$ as they contribute only radiatively, which would lead to sign changes when Eq.~\eqref{eq:1} is used as input. However as previously stated we are not considering this due to the strong constraints on these WCs from other processes.  
It should be noted that all logarithms are suppressed by the UV scale $\sim\Lambda^{-2}\log(m^2/\mu^2)$, which as larger scales $\mu\sim\Lambda$ are considered, decreases slower than the tree level $\Lambda^{-2}$ contributions.

\begin{figure}[!t]
	\includegraphics[width=0.45\textwidth]{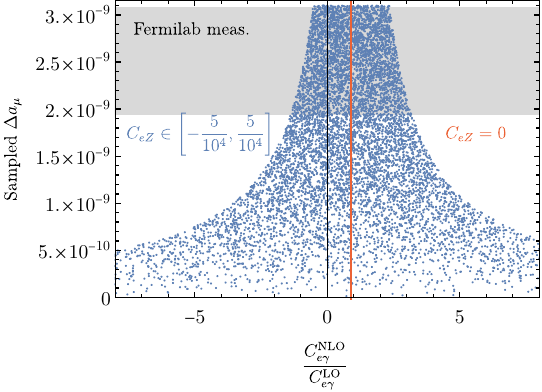}
	\caption{Impact of $C_{e Z}$ on $C_{e \gamma}$ for $\mu = m_\mu$. Relatively small values of $C_{e Z}$ can cause cancellations rendering the one-loop contribution of $C_{e \gamma}$ negligible and reproducing the Fermilab $\Delta a_\mu$. For larger sampled values of $C_{e Z}$ the $C_{e \gamma}$ K-factor increases to keep consistency with the Fermilab measurement.\label{fig:amuvskfactor}}
\end{figure}

We evaluate the importance of the one-loop contribution to $\Delta a_\mu$, when compared to the tree level, with the scale choice of the muon mass. To do so we consider the orthogonal rotation of the dipole operators,
\begin{equation}
	\begin{split}
		C_{e \gamma} &= c_W C_{e B} - s_W C_{e W}\;,\\
		C_{e Z} &= s_W C_{e B} + c_W C_{e W}\;,\\
	\end{split}
\end{equation}
such that at tree level only $C_{e \gamma}$ contributes to the muon's dipole moments. Taking into account combinations of $C_{e \gamma}$ and an additional operator, it is possible to evaluate how sizeable the correction factor from loop corrections (the K-factor) can be when $\Delta a_\mu$ lies within the allowed interval of Fermilab. We define this factor as the ratio of $C_{e \gamma}$ at NLO required to reproduce a value of $\Delta a_\mu$ divided by the value considering only tree level effects.
When no additional operators are present, the $C_{e \gamma}$ K-factor lies at $\sim 0.9$ for $\mu = m_\mu$ and increases logarithmically as $\mu \to \Lambda = 1$~TeV up to $\sim 1.2$.  
The dependence of the $C_{e \gamma}$ interval consistent with the Fermilab result is shown in Fig.~\ref{fig:mudepcegamma}. This shows that also in EFT (mirroring the findings in concrete UV models~\cite{Athron:2021iuf}), the anomalous magnetic moment measurement is linked to a sizeable coupling. While this is not directly visible from the tree level result, the size of the radiative correction (which is much larger than typical electroweak corrections) indicates a relative strong coupling of the EFT when the dipole operators alone are considered. The presence of additional SMEFT operators, however, can reproduce the muon $g-2$ anomaly without the need of such large radiative corrections. For example, sampling values of $\Delta a_\mu$ and $C_{e Z}$ we can see that even if $C_{e Z}$ is assumed to be of order $10^{-4}$, the anomalous magnetic moment can be reproduced with small radiative corrections as shown in Fig.~\ref{fig:amuvskfactor}.

\begin{figure*}[t!]
\subfigure[]{\includegraphics[width=8.3cm]{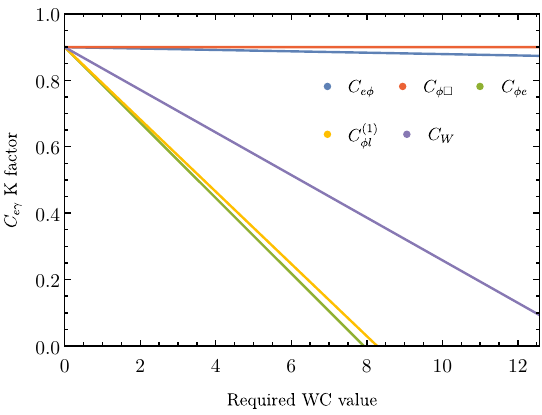}}
\hfill
\subfigure[]{\includegraphics[width=8.3cm]{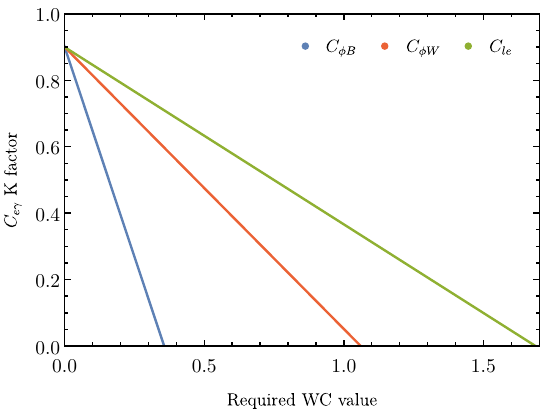}}
	\caption{Plots showing the minimum $C_{e \gamma}$ K-factor possible that can reproduce the anomalous magnetic moment depending on the range where an additional operator is sampled.  \label{fig:cegammakfactors}}
\end{figure*}

We can repeat this procedure for the rest of the SMEFT operators, calculating the minimum possible $C_{e \gamma}$ K-factor that is required to obtain consistency between the sampled $\Delta a_\mu$  and the Fermilab measurement, shown in Fig.~\ref{fig:cegammakfactors}. Phrased differently, we calculate how large a second SMEFT contribution needs to be in order to cancel the one-loop contribution of $C_{e \gamma}$ rendering the tree level as the only relevant contribution. In general, the radiative correction from the dipole operator remains relevant for most operators unless they approach the non-perturbative limit. Exceptions are the $C_{\phi B}$, $C_{\phi W}$ and $C_{l e}$ operators which, without acquiring sizeable values, can introduce cancellations in the NLO part rendering the loop order negligible. 

Due to its link to a direction in the SMEFT parameters space, $\Delta a_\mu$ is a scheme- and scale-dependent parameter, and should therefore be approached with the necessary caution as scale dependencies can be modified in actual scattering cross section calculations involving $C_{e \gamma}$. Being an observable, the experimental measurement of $g-2$ will be unaffected by unphysical contributions related to scheme choices, which should cancel when all relevant contributions are taken into account. The main result, however, remains that the bounds on the dipole operators are extremely tight, raising the question whether similar precision can be achieved through alternative channels. We will revisit this in Sec.~\ref{sec:tension}.

\subsection{$ a_\mu$ as a radiative SMEFT effect}
\label{subsec:finops}
The muon $g-2$ measurement can also be interpreted as an effect of BSM physics from operators that contribute at one loop level without leading to UV divergences for the dimension-six truncation.  When considering operators individually, $C_{\phi \square}$, $C_{e \phi}$ and $C_W$
cannot produce enough pull to lift the $g-2$ up to the Fermilab measurement unless their absolute value divided by $\Lambda^2$ exceeds the limit of $4 \pi/\text{TeV}^2$. For the electroweak scale of $246$~GeV this would correspond to a WC value of order one. 
Electroweak Precision Observables (EWPO) significantly restrict the allowed values of $C_{\phi l}^{(1)}$, $C_{\phi l}^{(3)}$ and $C_{\phi e}$~\cite{Dawson:2019clf}. The only remaining currently unconstrained operator that can provide an explanation of $g-2$ is four-fermion $C_{l e}$, which was also discussed in Refs.~\cite{Crivellin:2013hpa,Aebischer:2021uvt}. Lepton colliders will be able to place constraints on this operator~\cite{Bouzas:2021sif,deBlas:2022ofj}. 

\section{Avenues for tensioning the dipole operators}
\label{sec:tension}
The dipole operators $C_{e B}$ and $C_{e W}$ are difficult to constrain in collider environments. In this section, we explore a range of motivated avenues which could, a priori, provide bounds either with current experimental uncertainties or in the future. We consider the $Z\to \mu \mu$ decay, expected to be sensitive to $C_{e Z}$ as well as the muon-Higgs interactions. While the $\mu \mu \gamma$ and $\mu \mu Z$ vertices are sensitive to the dipole operators when the Higgs doublet is set to its vacuum expectation value, Higgs physics such as the 125 GeV Higgs boson's decay  $h \to \mu \mu$ is also sensitive to interactions that are linked to $a_\mu$. Additionally we include resolved photon emission $h\to \mu \mu \gamma$ which is directly sensitive to the $a_\mu$ operators at the cost of low statistical yield (yet with low experimental systematics as muons and photons are under very good control at the LHC). Looking towards the future, we consider the precision environment provided by the MUonE experiment~\cite{Abbiendi:2016xup} proposed at CERN, aiming to measure the hadronic contributions to the anomalous magnetic moment in $\mu e$ scattering, whose sensitivity can also be interpreted as bounds on the SMEFT operators considered in this work.

\subsection{$Z$ boson decay}
\label{sec:zdec}
The orthogonal rotation of the gauge fields $B_{\mu}$ and $W_\mu^3$ in the SM to the physical $Z$ and $\gamma$ also causes the appearance of the $C_{e B}$ and $C_{e W}$ operators in decays of the $Z$ boson to muons, along with additional BSM interactions. At dimension-six order only the linear contribution is relevant arising from interference with tree and virtual SM diagrams. For the discussion of the $Z$ decay we neglect loop contributions from SMEFT as they are unlikely to significantly impact the (relatively poor) constraints. 

Denoting as ${\cal{M}}_{\text{d6}}$ the tree level SMEFT amplitude and ${\cal{M}}_{\text{\text{SM}}}^{\text{tree}}$ (${\cal{M}}_{\text{SM}}^{\text{virt}}$) the tree (virtual) SM amplitude, the departure from the SM decay width is given by
\begin{equation}
	\delta \Gamma_{Z \to \mu \mu} = \frac{|\vec{p}_1|}{48 \pi m_Z^2} \bigg[2 {\cal{M}}_{\text{d6}}^* ({\cal{M}}_{\text{SM}}^{\text{tree}} +  {\cal{M}}_{\text{SM}}^{\text{virt}}) \bigg]\;,
\end{equation}
where we average over the initial polarizations of the $Z$ boson and integrate over the phase space ($\vec{p}_1$ denotes the three-momentum of one of the final states). The virtual SM amplitude terms are calculated at the renormalisation scale $\mu = m_Z$. 

The Particle Data Group (PDG)~\cite{ParticleDataGroup:2020ssz} reports an uncertainty of $0.18$~MeV on the decay width of $Z \to \mu \mu$ which we use to construct a $\chi^2$ and evaluate the allowed bounds at $68\%$ confidence level of each WC individually for comparability. Results are shown on Tab.~\ref{tab:zbounds} where the dipole operators are essentially unconstrained with bounds exceeding $4 \pi/\text{TeV}^2$. The muonic decay of the $Z$ boson can only efficiently constrain the three operators of the class $\psi^2 \phi^2 D$ that contribute to the anomalous magnetic moment. As with the EWPO that also constrain most of these operators, we see again that $C_{\phi e}$, $C_{\phi l}^{(1)}$ and $C_{\phi l}^{(2)}$ contributions cannot lift the muon $g-2$ up to the Fermilab/BNL findings without creating tension with other measurements.

\begin{table}[t!]
\begin{tabular}{c||c}
\toprule
	WC ($/ \Lambda^{2}$) & $68\%$ CL bound \\ 
\midrule
	$C_{\phi e}$ & $\left[-0.019, 0.019\right]$\\
	$C_{\phi l}^{(1)}$ & $\left[-0.016, 0.016\right]$\\
	$C_{\phi l}^{(3)}$ & $\left[-0.016, 0.016\right]$
\\ \bottomrule
\end{tabular}
	\caption{Bounds on WCs at $68$\% in units of $\text{TeV}^{-2}$ from the decay width of the $Z$ boson obtained by constructing a $\chi^2$ with the PDG error (see text). Contributions from dipole operators do not result in perturbatively meaningful constraints and are not shown.}
	\label{tab:zbounds}
\end{table}

\subsection{The $h \to \mu \mu \gamma$ channel}
\label{sec:gghmumua}
The decay of the Higgs boson to fermions was calculated in Ref.~\cite{Cullen:2020zof} (see also~\cite{Cullen:2019nnr} for more details on the calculation) which we utilise in order to identify the sensitivity to the dipole operators from the signal strength of $h \mu \mu$. The signal strength is can be expressed numerically as\footnote{We obtain the signal strength using the decay rates given in~Ref.~\cite{Cullen:2020zof}.}
\begin{equation}
	\mu = 1 + \frac{\text{TeV}^2}{\Lambda^2} (0.67 C_{e W} - 0.19 C_{e B} )\;,
\end{equation} 
at a scale $\mu = 125$~GeV. Bounds on the WCs are then obtained at $68$\% Confidence Level (CL) by constructing a $\chi^2 = (\mu - 1.19)^2/0.35^2$ using the PDG expectation value~\cite{ParticleDataGroup:2020ssz} which would correspond to $C_{e B}\in \left[-2.8, 0.8\right]$ and $C_{e W} \in \left[0.24, 0.81\right]$ using a UV scale $\Lambda = 1$~TeV. As the sensitivity is still orders of magnitude less than the anomalous magnetic moment, this motivates looking into alternative avenues.

\begin{figure}[!t]
	\includegraphics[scale=1]{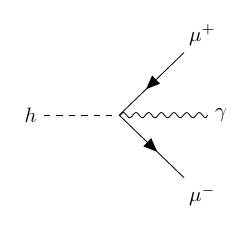}
	\caption{Vertex arising from the operators of Eq.~\eqref{eq:amuops} that contributes to the decay of the Higgs.\label{fig:diaghmumua}}
\end{figure}
The dependence of the ${\cal{O}}_{e W}$ and ${\cal{O}}_{e B}$ operators on the Higgs doublet suggests that $s$-channel processes with propagating Higgs bosons might provide additional tension on the bounds on $C_{e B}$ and $C_{e W}$ from the anomalous magnetic measurement. The operators generate the diagram shown in Fig.~\ref{fig:diaghmumua} which provides a final state at the LHC characterised by two muons and a hard photon, arising from a Higgs resonance (for a detailed discussion of $\mu \mu \to h \gamma$ process at a future muon collider see Ref.~\cite{Paradisi:2022vqp}). As the resonant Higgs is predominantly produced by gluon fusion at hadron colliders we consider the $p(g)p(g) \to h \to \mu^+ \mu^- \gamma$ channel. Background contamination arising from the SM will not be characterised by the same resonance structure which motivates a bump-hunt analysis around the Higgs peak. While the statistical yield is small, such data-driven analyses feature largely reduced systematic uncertainties similar to the Higgs boson's $h\to \gamma \gamma$ discovery mode. 

We use the {\sc{SMEFTsim}}~\cite{Brivio:2017btx,Brivio:2020onw} implementation of SMEFT\footnote{We use the effective $ggh$ interaction implemented in {\sc{SMEFTsim}} to model the Higgs production.} in the {\sc{UFO}}~\cite{Degrande:2011ua} format to generate events with {\sc{MadEvent}}~\cite{Alwall:2011uj,deAquino:2011ub,Alwall:2014hca}. In the final state we require at least two isolated leptons identified as muons with transverse momentum $p_T(\mu) > 15$~GeV in the central part of the detector (pseudorapidity must satisfy $|\eta(\mu)| < 2.5$). For an isolated muon, the sum of jet transverse momenta in the region $\Delta R = \sqrt{\Delta \phi^2 + \Delta \eta^2} < 0.4$ must be less than $50\%$ of the lepton's $p_T$,where $\phi$ denotes the azimuthal angle. Additionally at least one photon must be present with $p_T(\gamma) > 15$~GeV and $\eta(\gamma) < 2.5$ and along with the leading isolated muons, the Higgs mass $M_\text{reco}$ is reconstructed as the invariant mass of the four momenta total of the final states. An additional cut $110~\text{GeV} \leq M_\text{reco} \leq 135~\text{GeV}$ is imposed to select the region close to the resonant mass of the SM Higgs of $125$~GeV. 

\begin{figure}[t!]
	\includegraphics[width=0.45\textwidth]{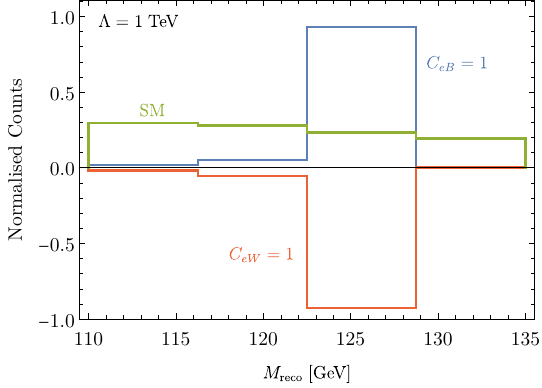}
	\caption{Example histogram for the reconstructed mass of the Higgs $M_\text{reco}$ showing the shape of the SM contribution and interference of effective interactions with the SM.\label{fig:cewceb_zdec}}
\end{figure}

\begin{figure}[t!]
	\includegraphics[width=0.45\textwidth]{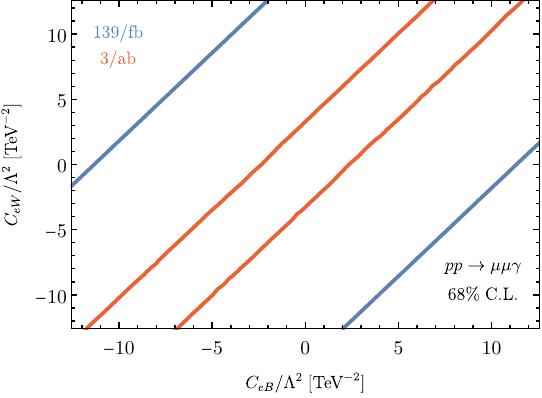}
	\caption{Exclusion contours in the $C_{e W}$-$C_{e B}$ plane from the $h \to \mu \mu \gamma$ channel for integrated luminosities of $139$/fb and $3$/ab. \label{fig:pptohmumua} }
\end{figure}

Differential cross sections in SMEFT can be expressed as
\begin{equation}
	{\text{d}} \sigma = {\text{d}} \sigma_{\text{\text{SM}}} + \frac{C_i}{\Lambda^2} {\text{d}}  \sigma_i \;,
	\label{eq:dsigmasmeft}
\end{equation}
where the first part arises from pure-SM interactions and the second term captures the interference of SM and dimension-six operators. We neglect any term suppressed by $\Lambda^{-4}$. Events are generated independently of the WC in this linearised setup and an example histogram for the differential distribution of the reconstructed Higgs mass is shown in Fig.~\ref{fig:cewceb_zdec}. A binned $\chi^2$ can be then constructed as the difference of the events with and without the EFT interactions squared, weighted by the SM statistical uncertainty. Performing a fit over the WCs of interest, results in the bounds shown in Fig.~\ref{fig:pptohmumua} for integrated luminosities of $139$/fb and $3$/ab. 

The sensitivity from this channel is once again severely limited compared to what is indicated from the anomalous magnetic moment. 

\subsection{MUonE}
\label{sec:muone} 
While we have considered the anomalous magnetic moment of the muon from the BSM perspective, there is still the possibility that higher-order hadronic contributions from the SM might resolve the $g-2$ anomaly. The dominant theoretical uncertainty of Eq.~\eqref{eq:1} arises from the hadronic vacuum polarisation which cannot be perturbatively computed at low energy scales as QCD becomes non-perturbative. Its determination, thus, relies on data-driven techniques through dispersive relations in the $e^+ e^- \to \text{hadrons}$ channel~\cite{Davier:2017zfy,Keshavarzi:2018mgv,Colangelo:2018mtw,Davier:2019can,Hoferichter:2019mqg,Keshavarzi:2019abf,Kurz:2014wya} which requires experimental results as input. A confirmation of the hadronic vacuum polarisation through first-principles lattice QCD techniques~\cite{Budapest-Marseille-Wuppertal:2017okr,RBC:2018dos,FermilabLattice:2019ugu,Gerardin:2019rua,Aubin:2019usy,Blum:2019ugy,Borsanyi:2020mff,Lehner:2020crt,Ce:2022kxy} would be convincing evidence that the Fermilab/BNL measurement indeed implies new physics. With Refs.~\cite{Budapest-Marseille-Wuppertal:2017okr,Ce:2022kxy} providing results for the hadronic effects that bring the $g-2$ closer to the Fermilab/BNL this still remains a highly relevant topic which needs to be resolved in the upcoming years. 

An alternative approach was proposed through the MUonE experiment~\cite{Abbiendi:2016xup}, aiming to determine the hadronic contributions with high accuracy. Using measurements of the differential cross section $\text{d}\sigma/\text{d}t$ in elastic $\mu e \to \mu e$ scattering, where $t$ is the (spacelike) squared momentum transfer, the hadronic contributions to the fine structure constant can be accurately determined. This can then be related to the respective contributions to the anomalous magnetic moment using a data-driven approach. Scattering off atomic electrons with a muon beam of $150$~GeV  from CERN, the experiment is expected to achieve an integrated luminosity of ${\cal{L}} = 1.5 \times 10^7$/nb for an expected cross section of $\sim 250~\mu{\text{b}}$. In such an experimentally well-controlled environment, theoretical uncertainties become limiting factors. There are ongoing efforts to improve predictions in the SM, see e.g.~\cite{Mastrolia:2017pfy,Gakh:2018sat,DiVita:2018nnh,CarloniCalame:2020yoz,Banerjee:2021mty,Budassi:2021twh}. In the following, we interpret 
deviations from the SM expectation of the MUonE experiment as new physics contributions (see also Ref.~\cite{Dev:2020drf}) to obtain a qualitative estimate of the experiment's sensitivity in light of our previous discussion. The impact of heavy particles is suppressed by their mass due to the relatively low centre-of-mass energy~\cite{Atkinson:2022qnl}, unless they are strongly coupled to the SM. We consider tree level contributions affecting the MUonE measurement from SMEFT in order to determine how the experiment can constrain the allowed ranges of relevant WCs, neglecting flavour-violating contributions from off-diagonal dimension-six operators.

\begin{figure}[t!]
	\includegraphics[width=0.45\textwidth]{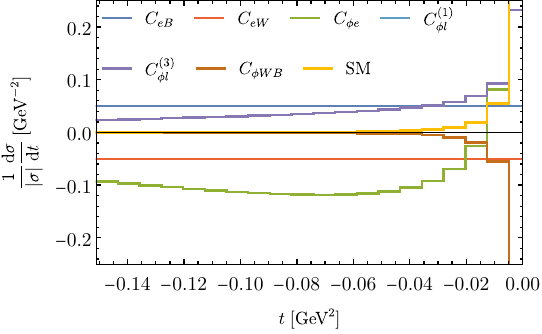}
	\caption{Histograms showing the shapes of the $\text{d} \sigma /\text{d} t$ distributions for the contributing WCs and the tree level SM expectation. For each line the WC is set to unity and the rest to zero.} 
	\label{fig:muonedsigdt}
\end{figure}

As before, the $\text{d}\sigma/\text{d}t$ distribution can be written linearly in terms of higher dimension operators as in Eq.~\eqref{eq:dsigmasmeft}, truncated at order $\Lambda^{-2}$. We calculate the unpolarized $\text{d} \sigma / \text{d} t$ differential distribution for $\mu^- e^- \to \mu^- e^-$ for the SM and for the interference of dimension-six operators with the SM (see Fig.~\ref{fig:muonedsigdt}). Bounds on the WCs are obtained with a binned 
\begin{equation}
\chi^2 = \sum_i {(\delta N_i)^2 \over \sigma_i^2}\;,
\end{equation} 
where $\delta N_i$ denotes the deviation in event counts from the SM $N^{\text{\text{SM}}}_i$ and $\sigma_i = \sqrt{\sigma_{\text{stat},i}^2 + \sigma_{\text{syst},i}^2}$ is the combination of statistical and systematic uncertainties. We use $\sigma_{\text{stat},i} = \sqrt{N^{\text{\text{SM}}}_i}$ and $\sigma_{\text{syst},i} = 10^{-5} N^{\text{\text{SM}}}_i$ which is the target systematic uncertainty by MUonE~\cite{Abbiendi:2016xup} (see also Ref.~\cite{Dev:2020drf}). 

We show the bounds on the relevant WCs in Tab.~\ref{tab:muonebounds}. The operators $C_{\phi e}$, $C_{\phi l}^{(1)}$ and $C_{\phi l}^{(3)}$ enter our calculations but their bounds exceed $4 \pi/\text{TeV}^2$; hence we have not included them in Tab.~\ref{tab:muonebounds}. However, aside from the measurement by Fermilab/BNL, the MUonE experiment can place the most stringent bounds from the approaches we consider in this paper but as in all the other scenarios these limits are orders of magnitude less restrictive than the muon $g-2$ result.

\begin{table}[t!]
\begin{tabular}{c||c}
\toprule
	WC (/$\Lambda^2$) & $68\%$ CL $[1/\text{TeV}^2]$  \\ 
\midrule
	$C_{e B}$ & $\left[-0.21, 0.21\right]$\\
	$C_{e W}$ & $\left[-0.39, 0.39\right]$
\\ \bottomrule
\end{tabular}
	\caption{Bounds on WCs at 68\% CL obtained with a $\chi^2$ analysis on the $d \sigma / dt$ distribution from MUonE. These correspond to $\abs{C_{e \gamma}} \leq 0.18$ and $\abs{C_{e Z}} < 12.75$.While operators of the $\psi^2 \phi^2 D$ class also contribute to the process, MUonE does not show significant sensitivity.	}
	\label{tab:muonebounds}
\end{table}

\section{Conclusions}
\label{sec:conc}
Having calculated the anomalous magnetic moment at one-loop order in SMEFT we observe that the higher-order contributions from the dipole operators $C_{e B}$ and $C_{e W}$ are expected to be sizeable. This supports the general finding for concrete UV extensions that new states need to be relatively strongly coupled to the muon when they are heavy (which is an underlying assumption of the EFT approach). Cancellations can arise from the virtual presence of additional operators, but they are required to be sizeable in most cases with the exceptions of $C_{\phi B}$, $C_{\phi W}$ and $C_{l e}$. 
When the operators are considered on a one-by-one basis, a possible explanation for the measured value of the muon $g-2$ from Fermilab can only arise from the dipole operators and the four-lepton interaction quantified by $C_{l e}$. This agrees with previous conclusions of Refs.~\cite{Crivellin:2013hpa,Aebischer:2021uvt} (our final $\Delta a_\mu$ expression is also in agreement). However, this leaves little room for a description of $g-2$ arising from new physics at the high-energy scales without introducing flavour-violating contributions (this path is explored in Refs.~\cite{Crivellin:2013hpa,Aebischer:2021uvt}). 

Under the conservative assumption that new physics appears specifically in relation to the muon, we explore the impact of the dipole operators in different experimental scenarios to evaluate whether it is possible to reach a similar level of precision as the Fermilab/BNL measurement. While the dipole operators contribute to the $Z$ decay to muons, the sensitivity of the channel is poor without any prospect for the dipole operators (though it can provide constraints on $C_{\phi e}$, $C_{\phi l}^{(1)}$ and $C_{\phi l}^{(3)}$). The presence of the Higgs doublet in the dipole operators allows us to additionally assess whether additional tension can be obtained from interactions of muons with the physical Higgs scalar. The $h \to \mu \mu \gamma$ channel receives SMEFT contributions from $C_{e B}$ and $C_{e W}$ at tree level and would therefore be an ideal channel to set bounds directly on $a_\mu$-related interactions. Statistical limitations do not render this mode competitive at the High-Luminosity LHC luminosity of $3$/ab. A precise determination of the $h\mu\mu$ signal strength can provide improved bounds due to the appearance of the dipole operators at one-loop level, yet not sensitive enough to address the Fermilab/BNL tension with the SM. 

Our last consideration regarding the dipole operators is the future MUonE experiment attempting to measure the hadronic contributions in the $g-2$ anomaly arising from the SM. When reinterpreted in terms of SMEFT interactions, this mode provides a competitive bound on the dipole operator compared to the other modes considered. However, these are far from the extreme precision provided by the targeted measurement of the anomalous magnetic moment at Fermilab/BNL. 

\bigskip
\noindent{\bf{Acknowledgements}} ---
We thank Anke Biek\"{o}tter, Christine T.H. Davies and Thomas Teubner for helpful discussions. A.B. and C.E. are supported by the STFC under grant ST/T000945/1. C.E. is supported by the Leverhulme Trust under grant RPG-2021-031 and the IPPP Associateship Scheme. P.S. is funded by an STFC studentship under grant ST/T506102/1.
\bibliography{paper.bbl} 

\end{document}